\documentclass[]{aa}

\newcommand{\mnras}[1]{MNRAS}
\newcommand{\apj}[1]{ApJ}
\newcommand{\apjl}[1]{ApJL}
\newcommand{\nat}[1]{Nature}
\newcommand{\aap}[1]{A\&A}
\newcommand{\aj}[1]{AJ}
\newcommand{\aaps}[1]{A\&AS}

\usepackage{epsf}
\usepackage{graphics}

\begin{document}

\title{CLASS B0827+525: `Dark lens' or binary radio-loud quasar?}

\author{L.V.E. Koopmans\inst{1,}\inst{5}, A.G.~de Bruyn\inst{2,}\inst{1},
  C.D.~Fassnacht\inst{3,}\inst{4}, D.R.~Marlow\inst{5,}\inst{6},
  D.~Rusin\inst{6}, R.D.~Blandford\inst{3}, I.W.A.~Browne\inst{5},
  P.~Helbig\inst{5,}\inst{1}, N.~Jackson\inst{5},
  S.T.~Myers\inst{6,}\inst{4}, T.J.~Pearson\inst{3},
  A.C.S.~Readhead\inst{3}, P.N.~Wilkinson\inst{5},
  E.~Xanthopoulos\inst{5}, H.~Hoekstra\inst{1}}

\mail{leon@jb.man.ac.uk} 
\offprints{L.V.E. Koopmans}

\institute{Kapteyn Astronomical Institute, P.O. Box 800, 9700 AV
  Groningen, The Netherlands \and NFRA, P.O.  Box 2, 7990 AA
  Dwingeloo, The Netherlands \and California Institute of Technology,
  Pasadena, CA 91125, USA \and NRAO, P.O.Box O, Socorro, NM 87801 \and
  University of Manchester, NRAL Jodrell Bank, Macclesfield, Cheshire
  SK11 9DL, England \and Department of Physics, University of
  Pennsylvania, Philadelphia, PA 19104, USA }

\authorrunning{Koopmans et al.}  \titlerunning{B0827+525: `Dark
lens' or binary radio-loud quasar?}

\thesaurus{12.07.1; 12.04.1; 11.17.4 B0827+525}

\date{Received ; accepted }

\maketitle

\begin{abstract}

We present radio, optical, near-infrared and spectroscopic
observations of the source B0827+525. We consider this source as the
best candidate from the {\sl Cosmic Lens All-Sky Survey} (CLASS) for a
`dark lens' system or binary radio-loud quasar.  The system consists
of two radio components with somewhat different spectral indices,
separated by 2.815 arcsec. VLBA observations show that each component
has substructure on a scale of a few mas. A deep {\it K}--band
exposure with the W.M.~Keck--II Telescope reveals emission near both
radio components. The {\it K}--band emission of the weaker radio
component appears extended, whereas the emission from the brighter
radio component is consistent with a point source. {\sl Hubble Space
Telescope} {\it F160W}--band observations with the NICMOS instrument
confirms this. A redshift of 2.064 is found for the brighter
component, using the LRIS instrument on the W.M.~Keck--II Telescope.
The probability that B0827+525 consists of two unrelated compact
flat-spectrum radio sources is $\sim$3\%, although the presence of
similar substructure in both component might reduce this.  

We discuss two scenarios to explain this system: (i) CLASS B0827+525
is a `dark lens' system or (ii) B0827+525 is a binary radio-loud
quasar.  B0827+525 has met {\sl all} criteria that thus far have in
100\% of the cases confirmed a source as an indisputable gravitational
lens system. Despite this, no lens galaxy has been detected with
$m_{\rm F160W}$$\le$23 mag. Hence, we might have found the first
binary radio-loud quasar. At this moment, however, we feel that the
`dark lens' hypothesis cannot yet be fully excluded.

\keywords{Cosmology: gravitational lensing -- Cosmology: dark matter
-- quasars: individual: B0827+525}

\end{abstract}

\begin{figure*}[t!]
\center
\resizebox{14.0cm}{!}{\includegraphics{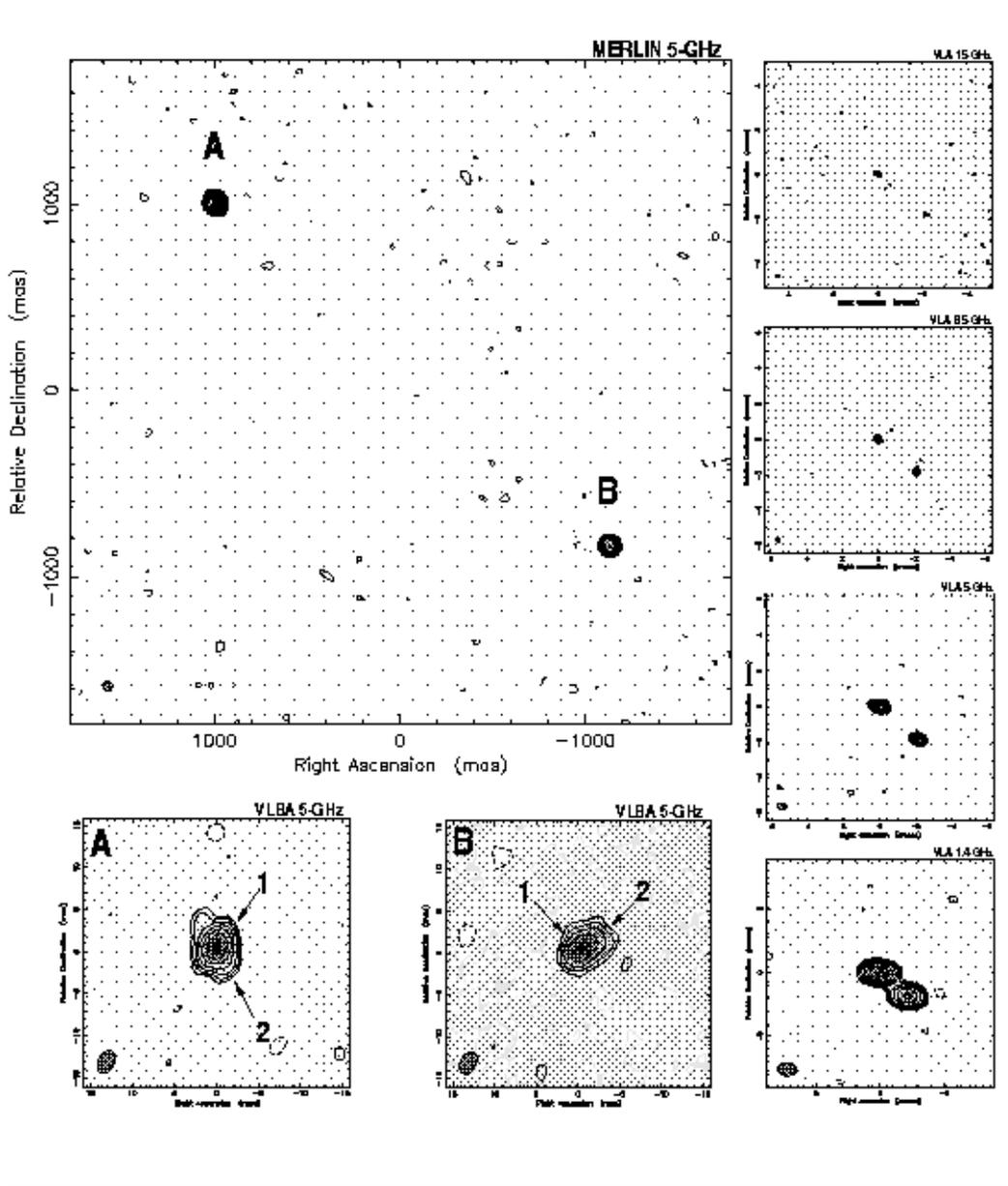}}
\vfill
\parbox[b]{\hsize}{
\caption{Summary of the VLA, MERLIN and VLBA radio images of B0827+525
(see text for more details). Contours start at the 3--$\sigma$ level and
increase by factors of 2. The first contours are 0.19 mJy~beam$^{-1}$,
0.37 mJy~beam$^{-1}$, 0.37 mJy~beam$^{-1}$, 1.15 mJy~beam$^{-1}$, 0.88
mJy~beam$^{-1}$, 0.40 mJy~beam$^{-1}$, 0.77 mJy~beam$^{-1}$, for the
MERLIN 5--GHz, VLBA 5--GHz (A), VLBA 5--GHz (B), VLA 15--GHz, 8.5--GHz
(discovery image), 5--GHZ and 1.4--GHz images, respectively}}
\end{figure*}

\section{Introduction}

There are several indirect ways to detect the presence of dark matter
in galaxies at relatively low redshifts (e.g. microlensing, rotation
curves of spiral galaxies, polar-ring galaxies, etc.). However, at
intermediate and high redshifts, weak and strong gravitational lensing
is the only method of detecting the presence of dark matter on galaxy
scales.

Gravitational lensing does not depend on the luminosity or color of
the lensing mass distribution. One can therefore expect to find
`dark-lens' galaxies in large gravitational-lens surveys, {\sl if}
they make up a significant fraction of massive galaxies (e.g. Hawkins
1997). On the basis of confirmed and candidate large-separation
($\ga$3 arcsec) gravitational-lens systems, which were all optically
selected, Hawkins (1997) concludes that around 75\% of all galaxies
could be `dark' (i.e. extremely underluminous). This high fraction was
shown to be inconsistent with observational results (Jackson et
al. 1998) from the {\sl Cosmic Lens All-Sky Survey} (CLASS; Browne et
al. 1999; Myers et al. 1999) and also very unlikely on the basis of a
number of statistical arguments (e.g. Kochanek, Falco \& Mu\~noz 1999;
Peng et al. 1999). A very high fraction of dark galaxies therefore
seems to be excluded, although a much smaller fraction ($\le$10\%)
cannot yet be ruled out. The detection of only a single dark-lens
galaxy would already have severe consequences for the standard picture
of galaxy formation (e.g. White \& Rees 1978). Such systems would have
allowed dark matter to collapse in structures massive enough to create
multiple images through gravitational lensing, but at the same time
prohibit baryonic matter to settle in the dark-matter halo and
initiate star formation.  Clearly, one has to be careful in dismissing
any candidate dark-lens galaxy as being either a binary quasar or a
chance alignment of unrelated quasars or AGNs, although undoubtedly
for the majority of candidates this will be the case.

In this paper, we present the strongest `dark-lens' candidate found in
the CLASS survey: CLASS B0827+525\footnote{Not to be confused with the
gravitational-lens system APM\,08279+5255 (Irwin et al. 1998).}.  In
Section 2, we present radio, optical, near-infrared and spectroscopic
observations of this system.  In Section 3, we derive constraints on
the lens-galaxy mass-to-light ratio and compare those with
observations of `luminous-lens' galaxies.  In Section 4, we summarize
our results and suggest future work on this system.

\section{Observations}

The CLASS survey aims to find all multiply-imaged flat-spectrum
($\alpha$$\le$0.5 for $S_\nu$$\propto$$\nu^{-\alpha}$) radio sources
in the northern hemisphere with a total flux density of $S_{\rm
5GHz}$$\ge$30\,mJy, a flux-density ratio between lens images $\le$10,
a Galactic latitude $|b|$$>$$10^\circ$ and a component separation
$\ge$0.3 arcsec.  The scientific goal of the survey is to create a
sample of GL systems, that can be used to study the structure and
evolution of lens galaxies at intermediate redshifts {\sl and}
constrain the cosmological parameters, in particular the Hubble
parameter (H$_0$). At present CLASS has discovered at least 17 new GL
systems (Browne et al. 1999; Myers et al. 1999).

\subsection{Radio observations}


If not specified otherwise, the radio data presented in this section
are flux-calibrated in the NRAO data-reduction package {\sf AIPS} and
self-calibrated, imaged and model-fitted in {\sf DIFMAP} (Shepherd
1997). To obtain flux-densities of the two radio components of
B0827+525, the uv-data was fitted by typically two Gaussian
components.

As part of the CLASS sample of around 15$\,$200 flat-spectrum radio
sources observed during the 1994, 1995 and 1998 {\sl
Very-Large-Array} (VLA) A--array seasons (Myers et al. 1999), a
30--sec snapshot of B0827+525 was made at 8.5\,GHz on 1995 August 13.
The calibrated VLA image (Fig.1) shows two unresolved components that
are separated by 2.8 arcsec and have a flux-density ratio of 2.7
[i.e. $S_{\rm A}$$\approx$24\,mJy and $S_{\rm B}$$\approx$9\,mJy].

With an average source density in the CLASS survey of $\approx$1 per
sq. deg, the probability of chance alignment of two unrelated compact
radio-loud quasars within $\le$2.8 arcsec is around 2$\cdot10^{-6}$,
whereas the lensing rate is around $10^{-3}$. In a sample of 15$\,$200
sources the occurrence of such a close alignment has a probability of
only 3\%. B0827+525 was therefore regarded as a candidate GL system.


To improve the resolution and obtain a two-point spectral index
between the radio components, a snapshot image of B0827+525 was made
on 1997 January 3 with the {\sl Multi-Element Radio Linked
Interferometer Network} (MERLIN) at 5~GHz. A flux-density ratio of
2.5 was found [i.e. S$_{\rm A}$$\approx$34 mJy and S$_{\rm
B}$$\approx$14 mJy] and a component separation of 2.8 arcsec, very
similar to the VLA 8.5--GHz observations.  Subsequent long-track
MERLIN 5--GHz observations were made on 1997 December 8, with a total
integration time on the source of $\approx$10 hours. The calibrated
image (Fig.1) has an rms-noise level of 60 $\mu$Jy\,beam$^{-1}$ and
a resolution of $\approx$50 mas. Both components remain
unresolved. The flux-densities are $S_{\rm A}$$\approx$34.0 mJy and
$S_{\rm B}$$\approx$14.8 mJy, respectively. No sign of extended
emission is detected in the image above a level of 0.6\%. The
similarity in flux-density ratio at 8.5 and 5\,GHz strengthened the
case for a lensing explanation of this system.


To improve the resolution by another order of magnitude, B0827+525 was
observed with the {\sl Very-Long-Baseline-Array} (VLBA) at 5\,GHz on
1997 August 2. Snapshots were obtained over a range of hour angles to
improve the uv-coverage. Phase referencing was done by rapid switching
between the nearby strong calibrator B0828+493 (2\,min) and B0827+525
(5--7\,min).  The total integration time on B0827+525 was 35 min.  The
final calibrated image (Fig.1) has a resolution of 2.5\,mas and an
rms-noise level of 0.1\,mJy\,beam$^{-1}$. The two radio components
were each modeled by two Gaussian sub-components.  Component A shows a
NE--SW extension, whereas component B shows an area of weak extended
emission in the NW--direction. The low SNRs of the NE--component in
image~A and the NW--extension of image~B does not warrant modeling
them by more than two subcomponents. The positions of the best-fit
Gaussians and their respective flux densities are listed in Table~1.

\begin{table*}[t!]
  \centering
  \begin{tabular}{lrrrr}
   \hline
         & A1 & A2 & B1 & B2 \\ 
        \hline
   $\Delta x$ (mas)     & 1064.02  & 1063.18  & $-$1074.16  & $-$1076.43  \\
   $\Delta y$ (mas)     &  930.74  &  929.89  &  $-$900.40  &  $-$899.78  \\ 
   maj. (mas)           &  --      & --       & 1.75      & 2.23 \\
   min/maj              &  --      & --       & 0.65      & 0.55 \\
   $\theta$ ($^\circ$)  & --       & --       & -31.40     & 5.80 \\ 
   S$_{4.9}$ (mJy)      & 24.8$\pm$0.1 & 6.8$\pm$0.1 & 9.3$\pm$0.1 & 3.6$\pm$0.1\\
  \hline
  \end{tabular}
\caption{VLBA 5--GHz data on B0827+525 (1997 Aug. 2) The VLBA
phase-center was RA 08$^{\rm h}$31$^{\rm m}$5\fs36200, DEC
52$^\circ$25$'$20\farcs24100. Positions ($\Delta x$, $\Delta y$) are
given with respect to to this phase-center. The 4.9--GHz flux-density
(S$_{4.9}$) of the sub-components and their deconvolved major axis,
axial ratio and position angle (N$\rightarrow$E) are also given. The
sub-components of component A are unresolved. The positional error is
2.5 mas divided by twice the SNR of the components. The errors on the
major and minor axes of the subcomponents in B are roughly 25\%}
\end{table*}


Both radio components show fairly flat spectra from 4.9 to 8.5 GHz.
The integrated flux density of B0827+525 (A+B) in the WENSS survey,
(70$\pm$5 mJy at 325 MHz, mean epoch 1992.2), is about equal to the
flux density at 1.4\,GHz in the NVSS, FIRST and WSRT observations
taken in the summer of 1997.  This suggests that the individual
component spectra must flatten to a spectral index of zero below
1.4\,GHz. Alternatively, one of the component spectra could become
inverted leaving room for the other to remain straight.  High
resolution low frequency data are needed to distinguish between these
possibilities.

A 2--month monitoring campaign (8 epochs) with the WSRT at 1.4 GHz
(with 15--arcsec resolution) in the period July-September 1997
indicates variability in the total flux density: the source flux
density slowly decreased from 71 to 58 mJy with a typical error of
about 1 mJy in individual measurements.  It is not clear to which
component these variations should be attributed.  The variation time
scale is rather short even though both components are compact. It is
perhaps more likely that the 1.4--GHz variations are due to refractive
interstellar scattering (RISS; e.g. Rickett et al. 1984), an
explanation requiring a less extreme size for the components.


\begin{figure}[t!]
\begin{center}
  \leavevmode
\vbox{%
  \epsfxsize=\columnwidth
  \epsffile{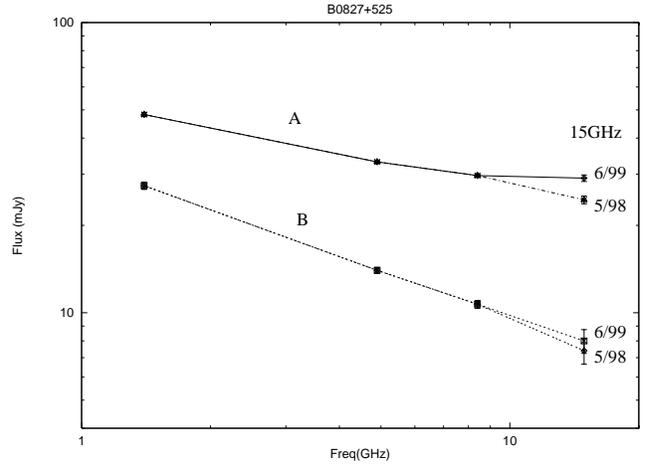}
}
\end{center}
\caption{Radio spectra of B0827+525 A and B, determined with the VLA
in A--array at 1.4, 4.9, 8.5 and 14.9 GHz on 1999 June 27}
\end{figure}

To examine the radio spectra of both components in more detail,
near-simultaneous VLA 1.4, 4.9, 8.5 and 14.9--GHz (L, C, X and
U--band) observations in A--array were done on 1999 June 27 (Table~2).
The radio spectra for both components are shown in Fig.2.  Components
A and B have fairly different radio spectra with spectral indices
$\alpha^{8.5}_{1.4}=0.27\pm0.01$ (A) and
$\alpha^{8.5}_{1.4}=0.52\pm0.01$ (B) (where
$S_{\nu}$$\propto$$\nu^{-\alpha}$). Also variability in component A at
14.9 GHz was detected from the 1998 May 15 and 1999 June 27 VLA
observations, which is more likely to be intrinsic and not RISS,
because the rms variability due to RISS decreases rapidly towards
higher frequencies (e.g. Rickett et al. 1995; Walker et al. 1998; see
also Koopmans \& de Bruyn 2000).

Variability at both lower (1.4\,GHz) and higher (15\,GHz) frequencies
makes a direct comparison of the spectral indices difficult.  The
formal difference in spectral index does not disqualify the source as
possible GL candidate, but complicates the argumentation in a `dark
lens' hypothesis. We will return to this issue in Sect.4.

\begin{figure*}[t!]
\begin{center}
  \leavevmode
\vbox{%
  \epsfxsize=\hsize
  \epsffile{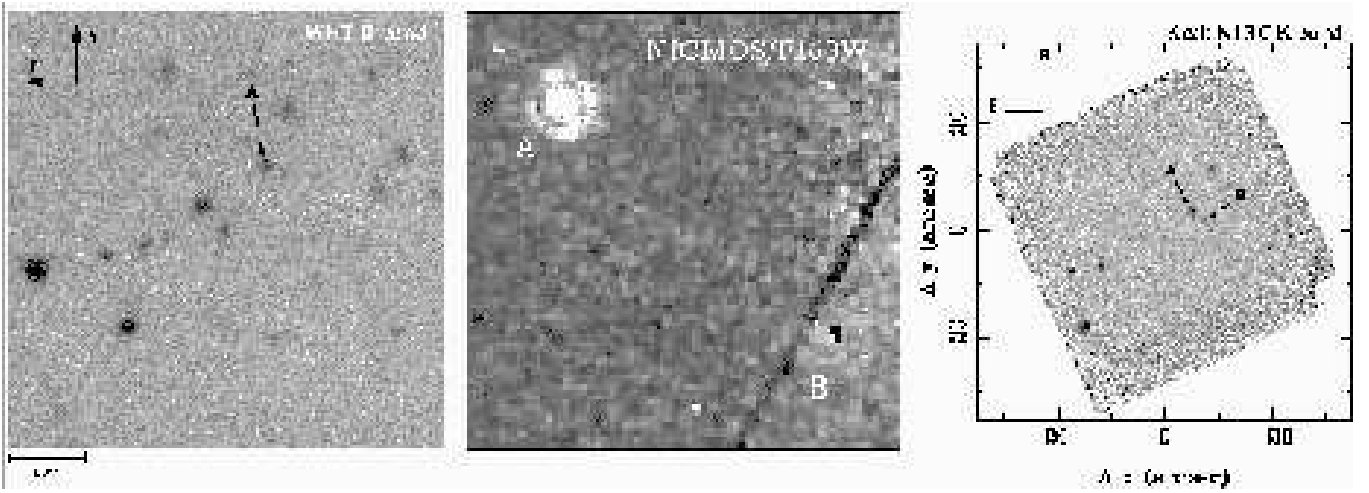}
}
\end{center}
\caption{{Left:} {\it R}--band image of B0827+525, observed 1998 May 19
with the WHT on La Palma. {Middle:} HST F160W NICMOS image of
B0827+525, observed on 1998 April 18.  The separation between
components A and B is 2.8 arcsec. {Right:} NIRC {\it K}--band image of
B0827+525, observed with the W.M~Keck--II Telescope on 1998 May 9}
\end{figure*}

\begin{table}[t!]
  \centering
  \begin{tabular}{crr}
   \hline
    $\nu$ (GHz) & $S_{\rm A}$ (mJy)  & $S_{\rm B}$ (mJy) \\
   \hline
   14.9  & 29.1$\pm$0.4 & 8.0$\pm$0.4  \\
   8.5   & 29.7$\pm$0.1 & 10.7$\pm$0.1 \\
   4.9   & 33.1$\pm$0.1 & 14.0$\pm$0.1 \\
   1.4   & 48.2$\pm$0.3 & 27.4$\pm$0.3 \\
  \hline
  \end{tabular}
  \caption{VLA multi-frequency data on B0827+525, taken
        near-simultaneously on 1999 June 27. The internal
        errors are indicated.}
\end{table}

\subsection{Optical and near-infrared observations}


{\sl Hubble Space Telescope} (HST) observations were made on 1998
April 18 with the {\sl Near-Infrared Camera and Multi-Object
Spectrometer} (NICMOS) at 1.6 $\mu$m (F160W), roughly corresponding to
ground-based {\it H}--band. The NIC1 camera was used which provides a
detector scale of 43 mas pixel$^{-1}$ and a field-of-view of
$11''$$\times$$11''$. The total exposure time was 2624 sec.  The data
were subjected to the standard NICMOS calibration pipeline in {\sf
IRAF}\footnotemark[1] which (i) corrects for known instrumental
effects (i.e. flat field and dark current), (ii) performs gain and
flux calibration and (iii) removes cosmic rays and flags degraded and
suspect data values\footnotemark[2]. The image was cut around
B0827+525 and rotated to correct to standard orientation.  The final
calibrated image is shown in Fig.3 (middle).  The primary northern
component (A) is unresolved and clearly detected ($m_{\rm
F160W}$=19.6$\pm$0.2 mag).  Component B is just barely detected as a
region of extended emission ($m_{\rm F160W}$=22.6$\pm$0.2 mag). No
sign of a lensing galaxy is seen in the image to $m_{\rm F160W}$=23
mag, although the extended emission near image B could partly be from
a possible lens galaxy. No optical correction (i.e. deconvolution) was
applied to the image, because the PSF distortion\footnotemark[2] is
not expected to change appreciably over the image separation of only
2.8 arcsec.

\footnotetext[1]{IRAF (Image Reduction and Analysis Facility) is
distributed by the National Optical Astronomy Observatories, which are
operated by the Association of Universities for Research in Astronomy
under cooperative agreement with the National Science Foundation.}

\footnotetext[2]{A detailed description of the NICMOS calibration
pipeline process, zero-point fluxes and PSF can be found in the Near
Infrared Camera and Multi-Object Spectrometer (NICMOS) handbook for
Cycle 10, available at the STScI web page.}


Similarly, B0827+525 was observed in {\it K}--band on 1998 May 9 with
the {\sl Near InfraRed Camera} (NIRC) at Keck. A total of 25 frames,
each having a 60--sec integration time, were obtained of the
field. The telescope was moved by approximately 10--15 arcsec between
individual integrations.  The seeing at the time of the observations
was 0.6--0.75 arcsec (FWHM). The exposures were combined and rotated
to correct the orientation and the resulting image is shown in Fig.3
(right). The observations confirm the conclusions drawn from the {\sl
HST} exposure, that image A is consistent with a point source and
image B seems somewhat extended.


An {\it R}--band image was taken on 1998 May 19 on the WHT. Two
600-sec exposures were obtained with a seeing of 1.4--1.5 arcsec
(FWHM). It was re-observed on 1998 May 20 with two exposures of
600--sec and one of 300--sec exposure, respectively. The seeing was
around 1.0 arcsec (FWHM). All exposures were combined and reduced in
the standard way in the data-reduction package {\sf IRAF}. The
resulting image (Fig.3, left) only show image A, whereas image B is
not detected above the noise level. Unfortunately no useful
calibration was obtained, but based on the exposure time and expected
noise-level, we roughly estimate for component A an apparent magnitude
of $m_{\rm R}$$\approx$22.0--22.5 mag and for component B a lower
limit $m_{\rm R}$$\ga$24.0 mag.

\subsection{Spectroscopic observations}

B0827+525 was observed with the {\sl Low Resolution Imaging
Spectrograph} (LRIS; Oke et al. 1995) on the W. M. Keck II Telescope
on the night of 1998 April 21. Three exposures were taken in longslit
mode with a total exposure time of 4500 sec. The 300 grooves mm$^{-1}$
grating was used, giving a dispersion of 2.44 \AA~per pixel.  The slit
was placed along both components, with a position angle of 50$^\circ$.
The final spectrum covers a wavelength range of 4024--9012 \AA.  The
spectra were reduced using standard {\sf IRAF} tasks.  A correction
for the response of the CCD was determined from observations of the
Oke standard star BD332642 (Oke 1990).  The corrected spectra were
weighted by the squares of their SNRs and co-added to produce the
final B0827+525 spectrum. Fig.4 shows the spectrum of component~A,
which has been smoothed using a boxcar kernel with a width of five
pixels. Emission lines associated with Si-IV $\lambda 1397$, C-IV
$\lambda1549$, He-II $\lambda1641$, and possibly C-III] $\lambda 1909$
are seen, establishing a source redshift of 2.064.  No emission at the
position of component B is detected. We also find no evidence in the
spectrum of component A for a second redshift, possibly from a lens
galaxy.

\begin{figure*}[t!]
\begin{center}
  \leavevmode
\vbox{%
  \epsfxsize=\hsize
  \epsffile{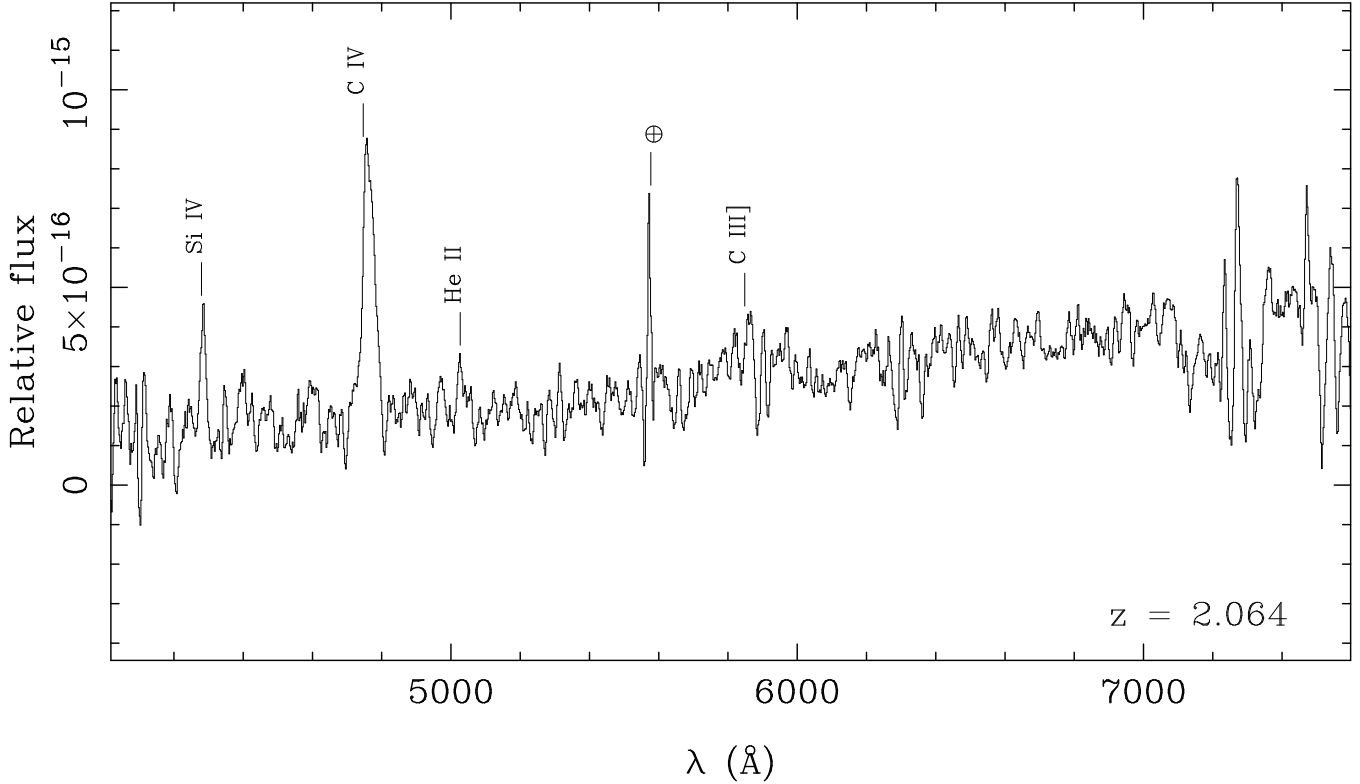}
}
\end{center}
\caption{LRIS spectrum of B0827+525, observed with the W.M~Keck~II Telescope
on 1998 April 21}
\end{figure*}

\section{Analysis}

Before proceeding with our analysis, let us summarize the most
important observational results. We have found that B0827+525 consists
of two radio components, separated by 2.8 arcsec. At least component~A
is variable, and both components have substructure at a scale of a few
mas.  The integrated radio spectra of both components from 1.4 to 14.9
GHz are somewhat different. The redshift for the brightest radio
component (A) is 2.064. Optical counterparts for both radio components
are found in {\it H} and {\it K}--band observations.  The emission
associated with the fainter radio component appears extended, which
could suggest that it is very faint galaxy emission.

\subsection{The `dark lens' hypothesis}

To compare the observed limits on the mass-to-light ratio of a
possible lens galaxy with those found for typical lens galaxies, we
first calculate the velocity dispersion of the lensing mass, using the
Singular-Isothermal-Sphere (SIS) mass model (Binney \& Tremaine
1987). This model relates the lens-image separation ($\Delta \theta$)
to a line-of-sight velocity dispersion ($\sigma$), given the
cosmological model and the source- and lens redshifts:
\begin{equation}
        \Delta \theta = 8\pi\cdot \left(\frac{\sigma}{c}\right)^2 \;
                \left(\frac{D_{\rm ds}}{D_{\rm s}}\right),
\end{equation}
where $c$ is the velocity of light, and $D_{\rm ds}$ and $D_{\rm s}$
are the angular-diameter distances between lens-source and
observer-source, respectively (e.g. Schneider, Ehlers and Falco
1992). The angular-diameter distance is a function of redshifts and
cosmological model.  Using the component separation, we can
furthermore calculate the mass contained inside the Einstein radius
\begin{equation}
     M(\Delta \theta)= 3.1\cdot 10^{10}\times
                \left(\frac{\Delta \theta}{1''}\right)^2 
                \left(\frac{D}{\rm Gpc}\right)\mbox{\rm~~M}_\odot,
\end{equation}
where $D$$\equiv$$D_{\rm d} D_{\rm s}/D_{\rm ds}$ and $D_{\rm d}$ is the
angular-diameter distance to the lens galaxy.
To calculate the luminosity of the lens galaxy, using the filter band
$\lambda$ (e.g. \textit{U, B, V,} $\dots$), we subsequently use the relation
\begin{equation}
        L_{\lambda}/{\rm L}_{\lambda, \odot} = 10^{ 
        {0.4\cdot (M_{\lambda,\odot}-m_\lambda+ {\rm DM+K}_{\lambda})}},
\end{equation}
\noindent
where DM is the distance modulus of the lens galaxy,
M$_{\lambda,\odot}$ is the absolute magnitude of the sun, $m_\lambda$
is the apparent magnitude of the lens galaxy and K$_{\lambda}$ is the
K--correction for the filter band $\lambda$. We use the lower limit of
$m_{\rm F160W}$$>$23 mag on the lens galaxy, found from the HST
exposure (Fig.3) to constrain the mass-to-light ratio in {\it
H}--band.  We use the {\it H}--band observations, because K
corrections are relatively small ($<$0.2 mag for $z$$\le$1 and $<$0.5
mag for $z$$\le$2; Poggianti 1997).  We do not use evolutionary
corrections and assume $M_{{\rm H},\odot}$=3.46 mag.

For a source redshift of 2.064, we have plotted the {\it H}--band
mass-to-light ratio in Fig.5 for two different cosmologies (A:
$\Omega_{\rm m}$=1 and $\Omega_{\Lambda}$=0; B: $\Omega_{\rm m}$=0.3
and $\Omega_{\Lambda}$=0.7) and H$_0$=65 km s$^{-1}$ Mpc$^{-1}$. We
assume a Friedmann-Robertson-Walker universe. A minimum mass-to-light
ratio around 100 is found, using cosmology B.  This mass-to-light
ratio implies a lens-galaxy redshift around 1.5 and a high velocity
dispersion of 400--500 km/s. For cosmology A, the minimum
mass-to-light ratio would increase to about 200.  From Jackson et
al. (1998) we find that the mass-to-light ratio in {\it H}--band for
B0827+525 is a factor 15--100 larger than the {\it H}--band
mass-to-light ratios of typical lens galaxies in the CLASS survey.

In Fig.6, we have plotted the expected {\it H}--band magnitude of the
lens galaxy as function of redshift, galaxy type and cosmology. We
used (i) the velocity dispersion determined from eqn.1, (ii) the
relation between velocity dispersion and {\it B}--band magnitude of
elliptical and spiral galaxies from Fukugita et al. (1991) and (iii)
the colors and K--corrections from Poggianti (1997). From Fig.6 an
upper limit on $m_{\rm H}$ of 17 mag is found, nearly independent of
galaxy type. If evolutionary corrections are applied (Poggianti 1997),
we find no significant differences in these results. The upper limit
is at least 6 mag brighter than the lower limit on the {\it H}--band
magnitude of the possible lens galaxy, whereas for all confirmed GL
systems (Jackson et al. 1998), for which the lens and source redshifts
are known, differences between the observations and the model are
$\la$1 mag.

Hence, if B0827+525 is a lens system then the object is indeed a `dark
lens' system. However, also image~B must be darkened and reddened by
the `dark lens' galaxy. For further discussion of this possibility we
refer to Sect.4.

\begin{figure}[t!]
\begin{center}
  \leavevmode
\hbox{%
  \epsfxsize=\columnwidth
  \epsffile{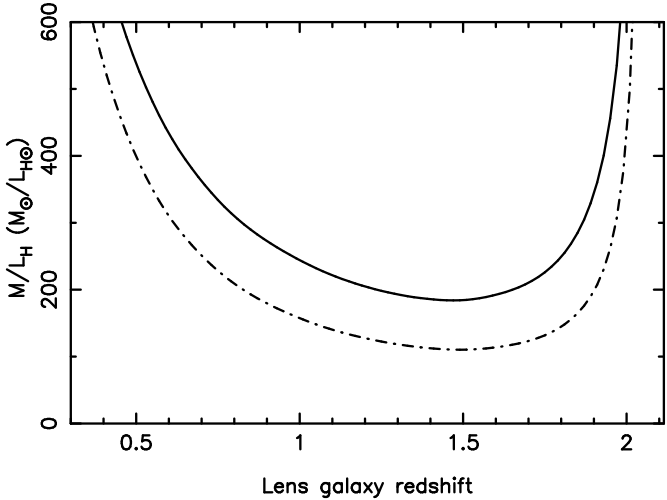}
}
\end{center}
\caption{Estimated lower limit on the {\it H}--band mass-to-light
ratio of the lens galaxy, as function of redshift (see text). The
solid line is for a ($\Omega_{\rm m}$=1, $\Omega_{\Lambda}$=0)
cosmology, the dot-dashed line for a ($\Omega_{\rm m}$=0.3,
$\Omega_{\Lambda}$=0.7) cosmology. We assume H$_0$=65 km s$^{-1}$
Mpc$^{-1}$}
\end{figure}

\subsection{The binary quasar hypothesis}

As argued in Sect.2.1 the probability of a chance alignment of two
radio-loud quasars is rather small. Hence, if the source is {\sl not}
a lens systems it is most likely a pair of physically close quasars.
We call this the `binary quasar' hypothesis even though we have no
evidence for the quasar nature of component B in the optical.

The compact flat-spectrum radio source population selected via CLASS
is generally believed to emit its radio emission in a highly
non-isotropic manner (e.g. Orr \& Browne 1982) with the radio axis
pointed in our direction within a small angle. As first argued by
Scheuer \& Readhead (1979) there should then be many radio-quiet
quasars (QSO) for every radio-loud quasar (QSR).  It is then obvious
that the probability of finding two related radio-loud quasars pointed
at us is rather small. In fact we can probably state that for every
QSR--QSR pair there should be around 20 QSR-QSO pairs in the CLASS
survey, if $\approx$5\% of all QSOs are also radio-loud,
i.e. $\ge$30\,mJy at 8.5\,GHz (e.g. Hooper et al. 1995; Bischof \&
Becker 1997).

Under the assumption that B0827+525 is the only QSR--QSR pair
(i.e. binary radio-loud quasar) in CLASS, the probability that a
flat-spectrum radio source is part of a QSR--QSR pair is around 1 in
15\,000. This means that the probability that a CLASS source is part
of a QSR--QSO pair is around 1 in 750. In the Large Bright Quasar
Survey (LBQS; Hewett et al. 1998) two QSO--QSO pairs were found from a
sample of around 1000 optically selected quasars, hence 1 in 500. This
number is very close to that found from the CLASS survey, which means
that the presence of one QSR--QSR pair in the CLASS survey is
consistent with the rate of QSO--QSO pairs in optical quasar surveys
(see also Kochanek et al. 1999).

However, if B0827+525 is a binary radio-loud quasar and we compare it
with the list of wide separation quasar pairs in Kochanek et
al. (1999), we notice two things: First, B0827+525, has the smallest
separation (2.8 arcsec or 23 kpc/$h_{50}$) of all quasar
pairs. Second, only 2 out of 13 non-lens quasar pairs have higher
redshifts (i.e. LBQS 1429-008 and Q2345+007). On the other hand,
B0827+525 would also be one of the largest separation lens systems
(e.g. Browne et al. 1999; Myers et al. 1999;
CASTLES survey\footnote{http://cfa-www.harvard.edu/glensdata/}). 
B0827+525 thus appears to be at the parameter-space border
delineated by the optical quasar pairs listed by Kochanek et
al. (1999) and the confirmed GL systems.  It is difficult to see how
significant these issues are in the context of binary quasars and how
it possibly relates to the fact that both quasars are radio-loud.

\section{Discussion \& Conclusions}

If taken at face value, most of the observational evidence seems to be
in favor of B0827+525 being a binary radio-loud quasar: (i) somewhat
different radio spectra from 1.4 to 14.9 GHz for the two components,
(ii) a different radio and optical brightness ratio, (iii) the weaker
radio component appears slightly resolved, (iv) the extended nature of
the optical and near-infrared emission near component B, whereas that
of image A is compact and (v) the agreement with the number statistics
of binary radio-loud quasars in the CLASS survey with those of
optical surveys.

\begin{figure*}[t!]
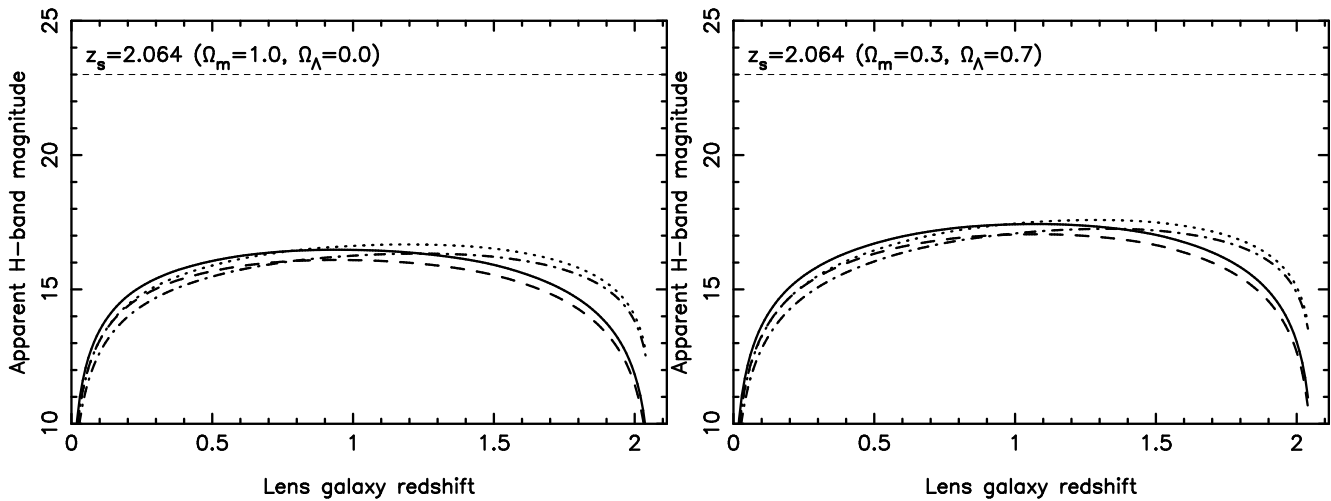

\begin{center}
  \leavevmode
\hbox{%
  \epsfxsize=\columnwidth
  \epsffile{fig6a_9501.epsi}
  \epsfxsize=\columnwidth
  \epsffile{fig6b_9501.epsi}
}
\end{center}
\caption{The estimated {\it H}--band magnitude of the lensing galaxy
for 4 galaxy types -- E (solid), S0 (dash), Sa (dot-dash) and Sc (dot)
--, plotted as function of redshift (see text). The horizontal dashed
line indicates the observed lower limit of 23 mag on the {\it H}--band
magnitude. Two cosmologies are shown}
\end{figure*}

However, the most interesting explanation of B0827+525, a ``dark
lens'' system, cannot convincingly be excluded and in view of the
cosmological importance of the presence of galaxy-sized concentrations
of dark matter -- even though they must be rare (e.g.  Jackson et
al. 1998) -- we now discuss why we believe the `dark lens' hypothesis
can not yet be discarded: (1) We have seen that at least one of the
radio components is variable. If they are lens images, a time delay
between them combined with variability of the source can result in a
difference in simultaneously measured spectral indices, especially if
the time delay is of the order of the variability time scale. (2) If
we place a massive `dark lens' galaxy near component B, containing a
large amount of dust, this galaxy would obscure most optical and a
significant fraction of the near-infrared emission from component
B. In the secure lens system B0218+357 a similarly large discrepancy
has been observed between the optical and radio brightness ratios of
the lens images (e.g. CASTLES survey), which can also be explained in
terms of obscuration by the high-column density of the rich ISM in the
lens galaxy (e.g. Wiklind \& Combes 1995; Menten \& Reid 1996; Combes
Wiklind \& Nakai 1997; Combes \& Wiklind 1997; Gerin et al. 1997;
Combes \& Wiklind 1998).  This dust, however, would probably not block
all of the lens-galaxy emission. In fact, most of the extended
emission seen near radio component B could be coming from the lens
galaxy itself and not from the quasar. (3) The optical brightness of
image B is surprisingly low for a radio-loud quasar, which might
suggest some form of extinction. (4) A high scattering measure,
associated with the ionized component of the ISM in the lens galaxy
(see for example Marlow et al. 1999; Jones et al. 1996), could
furthermore scatter-broaden radio component B, explaining why it
appears somewhat resolved.

So far in the `dark lens' hypothesis we have assumed that the lensing
mass is a single galaxy dominated by non-baryonic matter in its inner
parts and therefore extremely underluminous. If the lensing mass
distribution is a high-redshift ($z$$\ga$1) group or cluster, the
constraints on the mass-to-light ratio, which is usually an order of
magnitude larger in clusters and groups of galaxies than in the inner
parts of galaxies (e.g. van der Marel 1991; Carlberg, Yee \& Ellingson
1997), is somewhat alleviated. However, for most of the lens galaxies
from the CLASS survey, the {\it H}--band mass-to-light ratios are
around unity (Jackson et al. 1998). This is two orders of magnitude
smaller than the minimum {\it H}--band mass-to-light ratio found in
Sect.3.1 and can therefore not entirely account for the high
mass-to-light ratio within the context of a `normal' galaxy
evolution. Hence also in the case that the lensing mass is a group or
cluster, it must be underluminous. A possible candidate for this type
of mass concentration was recently found by Erben et al. (2000).

Thus, although the binary quasar hypothesis seems more likely at face
value, the `dark lens' hypothesis can not be ruled out
convincingly. However, the latter hypothesis does give concrete
predictions: First, the substructure in both radio components should
be related, although distorted by the intervening gravitational
potential.  We have obtained multi-frequency global VLBI data on
B0827+525 at three frequencies to test this. (2) If the variability of
the radio components is intrinsic, one should be able to correlate
their light-curves to find a time-delay.  This could definitively
prove or disprove whether B0827+525 is a `dark lens' system. (3) If
the extended {\it H} and {\it K}--band emission seen near component B
is partly from a lens galaxy, it should have a redshift smaller than
2.064. Near-infrared spectroscopy of this emission should therefore be
attempted.

\section*{Acknowledgments}

The authors would like to thank Lee Armus and David Hogg for obtaining
a Keck NIRC {\it K}--band image.  LVEK and AGdeB acknowledge the support
from an NWO program subsidy (grant number 781-76-101). This research
was supported in part by the European Commission, TMR Programme,
Research Network Contract ERBFMRXCT96-0034 `CERES'. The National Radio
Astronomy Observatory is a facility of the National Science Foundation
operated under cooperative agreement by Associated Universities, Inc.
The Westerbork Synthesis Radio Telescope (WSRT) is operated by the
Netherlands Foundation for Research in Astronomy (ASTRON) with the
financial support from the Netherlands Organization for Scientific
Research (NWO). MERLIN is a national UK facility operated by the
University of Manchester on behalf of PPARC. This research used
observations with the Hubble Space Telescope, obtained at the Space
Telescope Science Institute, which is operated by Associated
Universities for Research in Astronomy Inc.  under NASA contract
NAS5-26555.

\end{document}